\documentclass[pra,twocolumn,amsmath,amssymb,showpacs,floatfix]{revtex4}
\usepackage{graphicx}

\DeclareMathOperator{\argmin}{argmin}

\begin{document}

\title{Time evolution of Matrix Product States}

\date{\today}

\author{Juan Jos\'e Garc{\'\i}a-Ripoll}
\affiliation{Max-Planck-Institut f\"ur Quantenoptik,
  Hans-Kopfermann-Str. 1, Garching b. M\"unchen, D-85748, Germany}

\begin{abstract}
  In this work we develop several new simulation algorithms for 1D
  many-body quantum mechanical systems combining the Matrix Product
  State variational ansatz with Taylor, Pad\'e and Arnoldi
  approximations to the evolution operator. By comparing with previous
  techniques based on Trotter decompositions we demonstrate that the
  Arnoldi method is the best one, reaching extremely good accuracy
  with moderate resources. Finally we apply this algorithm to studying
  how correlations are transferred from the atomic to the molecular
  cloud when crossing a Feschbach resonance with two-species hard-core
  bosons in a 1D optical lattice.
\end{abstract}


\maketitle

The Density Matrix Renormalization Group (DMRG) method is a successful
technique for simulating large low-dimensional quantum mechanical
systems \cite{dmrg}. Developed for computing ground states of 1D
Hamiltonians, it is equivalent to a variational ansatz known as Matrix
Product States (MPS) \cite{mps}.  This relation has been recently
exploited to develop a much wider family of algorithms for simulating
quantum systems, including time evolution
\cite{vidal04,verstraete04b}, finite temperature
\cite{zwolak04,verstraete04b} and excitation spectra \cite{porras06}.
Some of these algorithms have been translated back to the DMRG
language \cite{dmrg-evol} using optimizations developed in that field
and introducing other techniques such as Runge-Kutta or Lanczos
approximations of the evolution operator
\cite{feiguin04,schmitteckert04}.

The MPS are a hierarchy of variational spaces, ${\cal S}_D$, [See
Eq.~(\ref{SD})] sorted by the size of its matrices, $D$. MPS can
efficiently represent many-body states of 1D systems, even when the
Hilbert space is so big that the coefficients of a pure state on an
arbitrary basis cannot be stored in any computer. While the accuracy
of this representation has been proven for ground
states \cite{verstraete05}, evolution of an arbitrary state changes the
entanglement among its parties, and a MPS description with moderate
resources (small $D$) might stop to be feasible.

We will take a pragmatic approach. First of all, MPS and DMRG
algorithms can compute truncation errors so that the accuracy of
simulations remains controlled. Second, we are interested in
simulating \textit{physically} small problems, such as the dynamics of
atoms and molecules in optical lattices. For such problems
small $D$ are sufficient to get a qualitative and even quantitatively
good description of the observables in our systems.  As we will see
below, the biggest problem is the accumulation of truncation errors
and not the potential accuracy of a given space ${\cal S}_D$ for
representing our states.

The outline of the paper is as follows. We first present many recently
developed simulation algorithms under a common formalism based on the
optimal truncation operator. We then introduce two new algorithms: one
of them is based on Pad\'e expansions of the evolution operator while
the other one uses linear combinations of MPS to increase the accuracy
(similar to an ``Arnoldi'' method). A comparison of all algorithms
using spin$-\tfrac{1}{2}$ models shows that all methods are strongly
limited by truncation and rounding errors and that the Arnoldi method
performs best for a given accuracy.

In the last part of the paper we apply these algorithms to study a
model of hard-core bosonic atoms going through a Feschbach resonance.
Current experiments \cite{molec-exp} with such systems have focused on
the number and stability of the formed molecules. In this work we
focus on the 1D many-body states and show that coherence is
transferred from the atomic component to the molecular one, so that
this procedure can be used to probe higher order correlations in the
atomic cloud.

\textit{Summary of algorithms}.-The space of MPS of size D is the set
of states of the form
\begin{equation}
  \label{SD}
  {\cal S}_D := \left\{
    \left(\mathrm{Tr}\prod_k A_k^{i_k}\right)|i_1\ldots i_N\rangle,~
    A_k^{i_n} \in \mathbb{C}^{D\times D}\right\}.
\end{equation}
Here $i_k=1\ldots d$ labels the physical state of the $k$-th lattice
site, the Hilbert space has a total of $d^N$ states and we sum over
repeated indices.

Since ${\cal S}_D$ is not a vector space, we cannot solve a
Schr\"odinger equation directly on it. Our goal is rather to
approximate the evolution at short times by a formula of the kind
$|\psi(t+\Delta t)\rangle \simeq {\cal P}_D \left[U_n(\Delta t)
  |\psi(t)\rangle\right].$ The operator ${\cal P}_D$ optimally
projects an arbitrary vector onto the manifold ${\cal S}_D,$ and
$U_n(\Delta t) = \exp(-i H \Delta t) + {\cal O}(\Delta t^n)$ is itself
an approximation to the evolution operator.

This formulation applies qualitatively to all recently developed MPS
and DMRG algorithms.  For instance,
Refs.~\cite{vidal04,verstraete04b,dmrg-evol} use a Trotter
decomposition of a nearest neighbor Hamiltonian $H=\sum_m H_{m,m+1}$
\begin{eqnarray}
  U_2(\Delta t) = e^{-i \sum_k H_{2k,2k+1}\Delta t/2}
  e^{-i \sum_k H_{2k+1,2k+2}\Delta t}&&\nonumber\\
 \times e^{-i \sum_k H_{2k,2k+1}\Delta t/2},
\end{eqnarray}
while the Runge-Kutta algorithm in Ref.~\cite{feiguin04} is
equivalent to a fourth order expansion of the exponential $U_4(\Delta
t) = \sum_{n=0}^{4} \tfrac{1}{n!} (-i\Delta t H)^n$.  In practice
there are differences between all cited implementations, such as
truncating in between applications of the exponentials or replacing
the previous Taylor expansion by a product of truncated first order
monomials, $\prod_k {\cal P}_D(1+\alpha_k H)$.

\begin{figure}[t]
  \includegraphics[width=\linewidth]{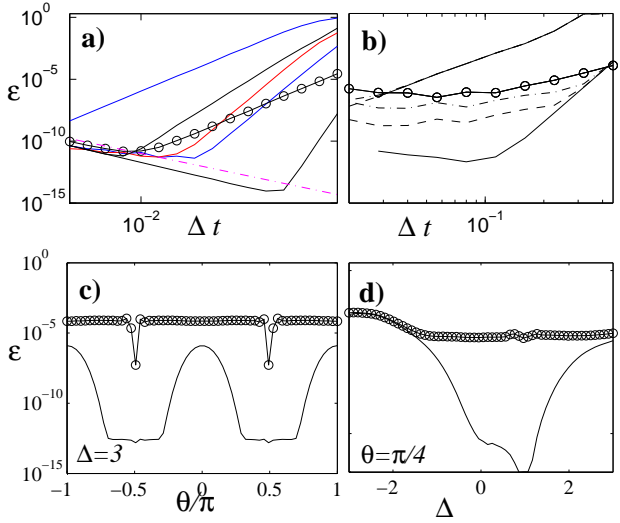}
  \caption{(a) Error, $\epsilon$, vs. time step, $\Delta t,$ for
    simulations with 8 spins, $D=16$ and $T=10$.  We compare Trotter
    expansions (circles), Runge-Kutta of 4th order (red), Pad\'e
    expansions of 2nd and 4th order (blue) and Arnoldi with 4 and 8
    vectors (black). In dash-dot line we plot the estimate
    $\varepsilon = (T/\Delta t)^2 10^{-16}$. (b) Same as plot but for
    $16$ spins and comparing only Arnoldi-4, Trotter and Arnoldi-8
    (top to bottom) for $D=64,80$ and $128$ (dash, dash-dot, solid).
    While (a-b) are done for $\theta:=\text{arctan}(B/J)=\pi/4$,
    $\Delta=0$, we have scanned the whole Hamiltonian space,
    $\theta\in[0,2\pi]$, $\Delta\in[-3,3]$. In (c-d) we show sections
    along $\theta$ and along $\Delta$, using 16 spins, $T=5$, $D=32$
    and the methods from (b).}
  \label{fig-exact}
\end{figure}

In this work we present two new methods. The first method uses a Pad\'e
approximation to the exponential
\begin{equation}
  \label{pade}
  U_n(\Delta t) = \frac{\sum_k \alpha_k H^k}{\sum_k \beta_k H^k},
\end{equation}
computed with same order polynomials in the numerator and denominator
\cite{moler03}.

The second method combines ideas from
Refs.~\cite{porras06,schmitteckert04}. First of all, it is possible to
prove that a linear combination of MPS, such as $\sum_{k=1}^{N_v}
c_k|\phi_D^{(k)}\rangle,$ resides in a space of bigger matrix product
states, ${\cal S}_{N_vD}$.  In the language of DMRG, $N_v$ vectors
each of size $D$ provide us with an effective basis of size $N_vD^2$.
This optimistic estimate is only possible when the vectors
$|\phi_D^{(k)}\rangle$ are linearly independent. Our choice will be
\begin{equation}
  |\phi_{k+1}\rangle \simeq {\cal P}_D\left(H|\phi_{k}\rangle -
    \sum_{j\leq k} \frac{\langle\phi_j|H|\phi_k\rangle}{\langle\phi_j|\phi_j\rangle}
    |\phi_j\rangle\right),
  \label{phikp1}
\end{equation}
with initial condition $|\phi_0\rangle := |\psi(0)\rangle.$ Finally,
defining the matrices $N_{ik} := \langle \phi_j |\phi_i\rangle$ and
$H_{ik} := \langle \phi_j | H |\phi_i\rangle$ we compute an Arnoldi
estimate of the exponential
\begin{equation}
  \label{Arnoldi}
  |\psi\rangle :=
  {\cal P}_D \sum_k [e^{-i\Delta tN^{-1}H}]_{k0} |\phi_k\rangle.
\end{equation}
This algorithm involves several types of errors. The error due to
using only $N_v$ basis vectors is proportional the norm of the vector
$\phi_{N_v}$ as in Lanczos algorithms \cite{lanczos}. Truncation
errors arising from ${\cal P}_D$ can be computed and controlled during
the numerical simulation. It is important to remark that while the
errors in Eq.~(\ref{phikp1}) may be compensated by adding more
vectors, the biggest errors arise from the final truncation in
Eq.~(\ref{Arnoldi}).

\textit{Implementation}.- All algorithms build on the same nonlinear
operator, ${\cal P}_D$, which optimally approximates a state using the
elements of ${\cal S}_D$. To truncate a linear combination of vectors
we use the definition \cite{verstraete04b}
\begin{equation}
  \label{PDk}
  {\cal P}_{D}\sum_k c_k |\phi^{(k)}\rangle
:= \underset{\psi\in{\cal S}_{D}}{\argmin}
\left\Vert |\psi\rangle - \sum_k c_k|\phi^{(k)}\rangle\right\Vert
\end{equation}
If we rather want to approximate the action of an operator that can be
decomposed as $U=X^{-1} Y$, we will apply a generalization of the
correction vector method \cite{dmrg}
\begin{equation}
  \label{PD}
  {\cal P}_{D}\left(X^{-1}Y | \phi \rangle \right)
  := \underset{\psi \in S_{D}}{\argmin}
  \Vert X|\psi\rangle - Y|\phi\rangle\Vert.
\end{equation}
This last formula has been used for the Pad\'e (\ref{pade}) equation
and for the Runge-Kutta methods, where $X$ and $Y$ are polynomials of
the Hamiltonian.

One may quickly devise a procedure for computing the
minimum of Eq.~(\ref{PD}) based on the definition of distance:
\begin{equation}
  \Vert X|\psi\rangle - Y|\phi\rangle\Vert^2
  =\langle\psi|X^\dagger X |\psi\rangle
  -2\Re\langle\psi|X^\dagger Y|\phi\rangle
  + \Vert{Y|\phi\rangle}\Vert^2.\label{distance}
\end{equation}
All scalar products in Eq.~(\ref{distance}) are quadratic forms with
respect to each of the matrices in the states $|\phi\rangle$ and
$|\psi\rangle$. The distance is minimized by optimizing these
quadratic forms site by site, or two sites at a time, sweeping over
all lattice sites until convergence to a small value which will be the
truncation error \cite{verstraete04b}.

The performance our algorithms is upper bounded by ${\cal O}(N_v N_H
D^6/L)$, where the $N_H$ is related to the number of operators in
$\{X, Y, X^\dagger X, Y^\dagger Y,\ldots\}$, $L$ is the size of the
problems and $N_v$ is the number of vectors in Eq.~(\ref{PDk}). For
open boundary conditions the cost reduces by ${\cal O}(D^2)$.

\textit{Comparison}.- We tested all algorithms by simulating the
evolution under the family of spin-$\tfrac{1}{2}$ Hamiltonians with
nearest-neighbor interactions
\begin{equation}
  \label{Hamiltonian}
  H =  \sum_k \left[ J(s^x_k s^x_{k+1} + s^y_k
    s^y_{k+1} + \Delta s^z_k s^z_{k+1})+ B s^z_k\right],
\end{equation}
of an initial state, $|\psi(0)\rangle \propto (|0\rangle +
|1\rangle)^{\otimes L},$ where $|0\rangle$ and $|1\rangle$ are the
eigenstates of $s^z$. By restricting ourselves to ``small'' problems
($L\leq 16$), we could compare all algorithms with accurate solutions
based on exact diagonalizations and the Lanczos algorithm
\cite{lanczos}.

\begin{figure}[t]
  \includegraphics[width=\linewidth]{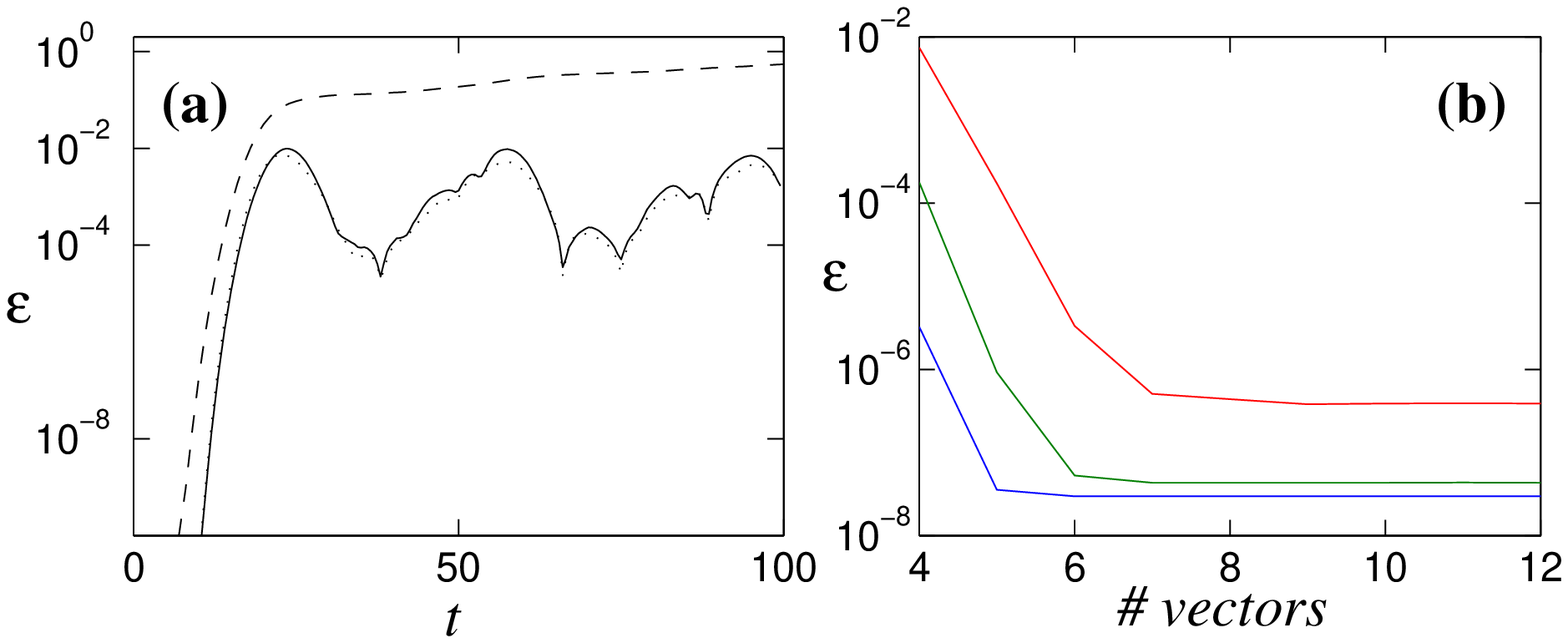}
  \caption{(a) Error made by approximating
    $\exp(-iHt)|\psi(0)\rangle$ with a vector in $S_{64}$ (solid)
    and error made when simulating this Schr\"odinger equation using
    the Arnoldi method with 8 vectors, $D=64$ and $\Delta t=0.16$
    (dashed).  (b) Similar as before, errors in the Arnoldi method for
    varying number of vectors, $T=10$ and $\Delta t= 0.04, 0.16, 0.32$
    (bottom to top).}
  \label{fig-truncation}
\end{figure}

In Fig.~\ref{fig-exact}a we plot the errors of the different methods,
$\varepsilon := \Vert \psi_D(T) - U(T)\psi(0)\Vert^2$, for 8 spins, $D
= 16$, $J=2B$ and $\Delta=0$. Since $S_{D}$ contains all possible
states, ${\cal P}_{D} = 1$ and we expect no truncation errors.
Therefore, for medium to long time steps the Trotter, Taylor and
Pad\'e approximations show the expected behavior, with an error
growing like ${\cal O}(\Delta t^2)$ or ${\cal O}(\Delta t^4)$. The
errors of the Arnoldi decompositions qualitatively follow ${\cal
  O}(\Delta t^{N_v})$.  However, all methods break these ideal laws at
some point, acquiring an error of order ${\cal O}(\Delta t^{-2})$,
which is exponential in the number of steps $T/\Delta t$.  This growth
signals the finite accuracy of the optimization algorithms, due to the
limited precision of the computer.  Roughly, since current computers
cannot compute the norms of vectors, $\Vert \psi\Vert^2$, with a
relative error better than $10^{-16}$, a worst case estimate is
$\epsilon = (T/\Delta t)^2 10^{-16}$ [See Fig.~\ref{fig-exact}a].

We have compared the best methods, that is the Trotter decomposition
and the Arnoldi methods with 4 and 8 vectors, using a bigger lattice
with 16 spins, and smaller matrices, $D=64,80$ and $128.$ As
Fig.~\ref{fig-exact}b shows, the $XY$ model establishes entanglement
along the lattice, the smaller matrices cannot account for this and
the errors grow exponentially.  Figure~\ref{fig-truncation}a shows how
the errors in the Arnoldi method are correlated to the errors made
when approximating the exact solution with a MPS of fixed size. One
could think that by increasing the number of vectors we should be able
to also increase the integration time-step and decrease the
truncations even for fixed $D$, but as Fig.~\ref{fig-truncation}b
shows, this not entirely true.

Summing up, one should use the method that allows for the longest time
steps and the least number of truncations (or applications of ${\cal
  P}_D$) and mathematical operations. All methods have an optimal time
step which is a compromise between the errors in $U_n$ and the
rounding and truncation errors made on each step.  Regarding
performance, the Arnoldi decomposition is competitive with the Trotter
decomposition as it reaches the same accuracy for longer time steps.
Furthermore, there is a huge potential for parallelizability, roughly
${\cal O}(N_vN_H/L)$, which all other presented algorithms lack.

The fact that previous results are model-independent is confirmed by a
systematic scanning of all possible Hamiltonians in
Eq.~(\ref{Hamiltonian}). A selection is shown in
Fig.~\ref{fig-exact}c-d. The Arnoldi method is shown to be the most
accurate one, even for gapless problems. When the Arnoldi method fails
it is due to truncation errors ---the evolved state cannot be
accurately represented by MPS---, which affect equally all MPS/DMRG
algorithms [Fig.~\ref{fig-exact}d].

\textit{Molecules in the lattice}.- We have used the Trotter and
Arnoldi methods to simulate the conversion of bosonic atoms into
molecules, when confined in an optical lattices and moving through a
Feschbach resonance \cite{molec-exp}. The goal is to study how
correlation properties are transferred from the atoms into the
molecules and how this dynamics is affected by atom motion and
conversion efficiency.

The effective model combines the soft-core Bose-Hubbard model used to
describe the Tonks gas experiments \cite{paredes04}, with a coupling to
a molecular state \cite{dickerscheid05}
\begin{eqnarray}
  H &=& -J \sum_{\langle i,j\rangle,\,\sigma} a^\dagger_{i\sigma} a_{j\sigma}
  + \sum_{i,\sigma,\sigma'}\frac{U_{\sigma,\sigma'}}{2}
  a^{\dagger}_{i\sigma} a^{\dagger}_{i\sigma} a_{i\sigma} a_{i\sigma}
  \label{Hubbard}
  \\
  &+& \sum_i \left\{ (E_m + U_m n^{(a)}_i) n^{(m)}_i +
    \Omega [b^\dagger_i a_{i\uparrow} a_{i\downarrow} +
    H.~c.]\right\}.
  \nonumber
\end{eqnarray}
Here, $a_{i\uparrow}$, $a_{i\downarrow}$ and $b_i$ are bosonic
operators for atoms in two internal states and the molecule; $n^{(a)}$
and $n^{(m)}_i$ are the total number of atoms and of molecules on each
site, and we have the usual two-level coupling with Rabi frequency
$\Omega$ and detuning $\Delta := E_m - U_m$.  For simplicity, we will
assume that atoms and molecules interact strongly among themselves
($U_{\uparrow,\uparrow},U_{\downarrow,\downarrow},U_{m} \to \infty$),
so that we can treat them as hard-core,
$a_{i,\sigma}^2,b_i^2,a_{i\sigma}b_i=0$. Also since molecules are
heavier, we have neglected their tunneling amplitude, although that
could be easily included.

\begin{figure}[t]
  \includegraphics[width=\linewidth]{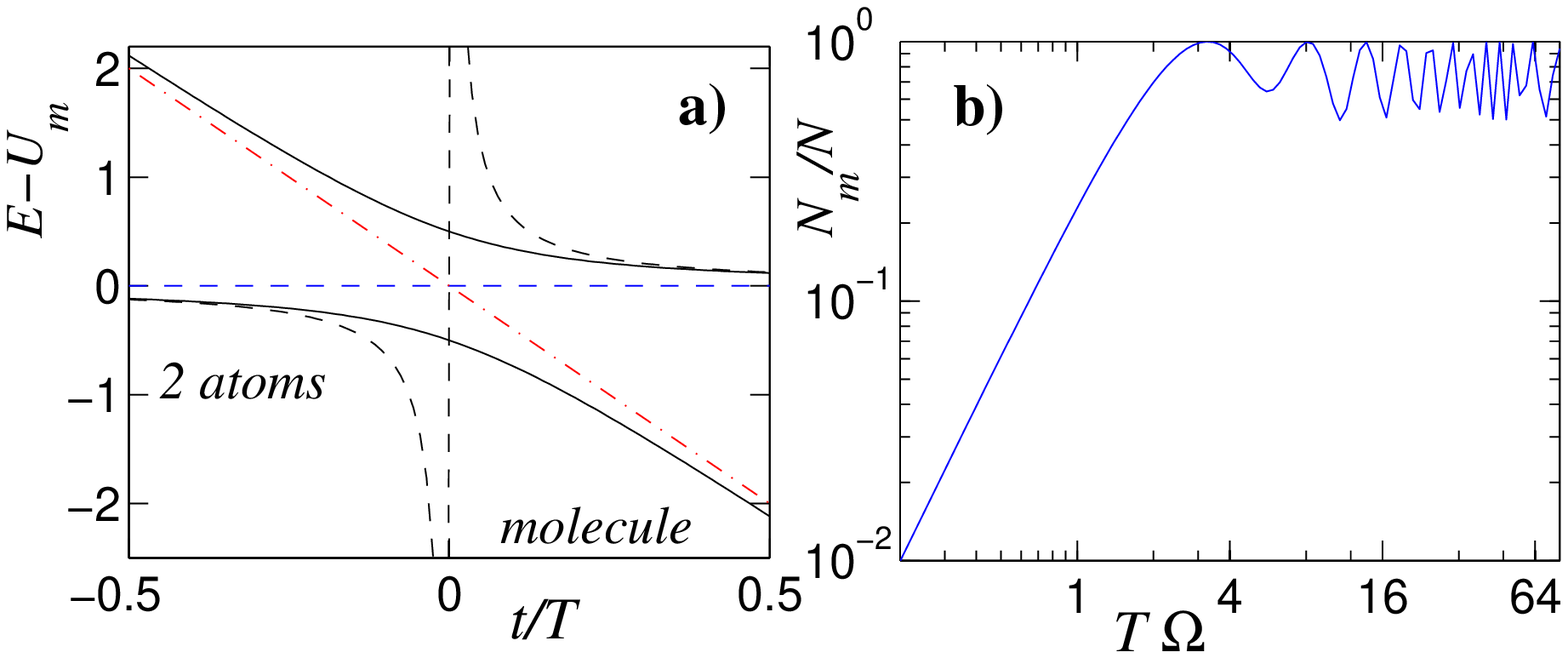}
  \caption{Dynamics of a site with two atoms when the energy of the
    molecules is ramped linearly: $E_m = U_{\uparrow\downarrow} +4
    \Omega (1 - 2t/T)$. We plot (a) the instantaneous energy levels
    (solid), and (b) the fraction of atoms converted into molecules.}
  \label{fig-molec-L2}
\end{figure}

As the energy of the molecular state is shifted from $E_m \gg U_m$
down to $E_m \ll U_m$, the ground state of Eq.~(\ref{Hubbard}) changes
from a pair of coupled of Tonks gases, to a purely molecular
insulator. We want to study the dynamics of this crossover as $E_m$ is
ramped slowly from one phase to the other.

The simplest situation corresponds to no hopping: isolated atoms
experience no dynamics, while sites with two atoms may produce a
molecule. The molecular and atomic correlations at the end of the
process are directly related to two-body correlations in the initial
state \cite{other-papers},
\begin{eqnarray}
  \langle m_k^\dagger m_k\rangle_{t=T} &\sim& \langle n_{k\uparrow}
  n_{k\downarrow}\rangle_{t=0},\label{corr}\\
  \langle a_k^\dagger a_k\rangle_{t=T} &\sim&
  \langle a_k^\dagger a_k\rangle_{t=0} - \langle n_{k\uparrow}
  n_{k\downarrow}\rangle_{t=0}.
\end{eqnarray}
Therefore, this process can thus be used as a tool to probe quantum
correlations between atoms. Studying the two-level system
$\{a^\dagger_{k\uparrow}a^{\dagger}_{k\downarrow}|0\rangle,
b_k^\dagger|0\rangle\}$, we conclude that for this process to work
with a $90\%$ efficiency, the ramping time should be larger than
$T\sim 1.5/\Omega$ [See Fig.~\ref{fig-molec-L2}(b)].

\begin{figure}[t]
  \includegraphics[width=\linewidth]{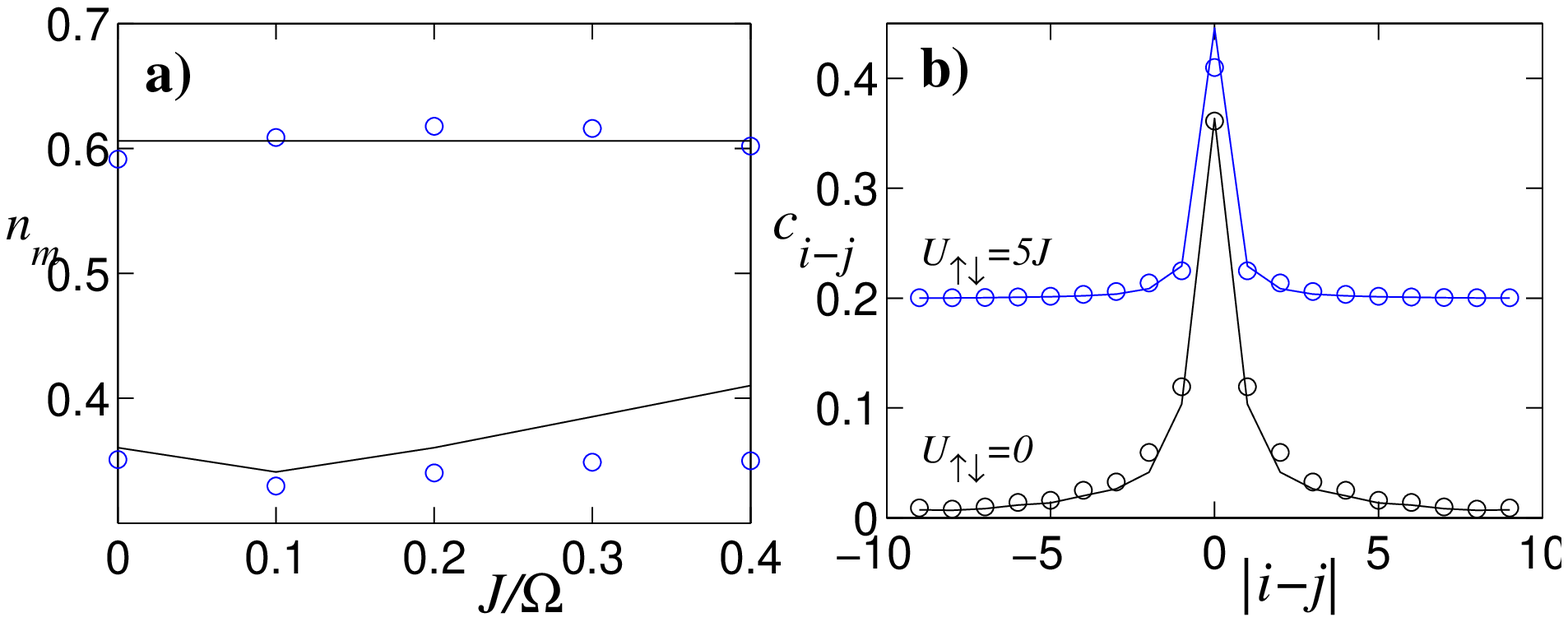}
  \caption{(a) Fraction of atoms converted into molecules vs.
    adimensionalized hopping amplitude, for
    $U_{\uparrow,\downarrow}=0$ and $U_{\uparrow\downarrow}=1$, from
    top to bottom. Circles show the outcome of the numerical
    experiment, while solid lines contain the ideal fraction
    (\ref{corr}). The case $J=0$ uses an initial condition
    $J=0.1\Omega$ and then switches off tunneling before ramping. (b)
    Correlations of the molecular state, $\langle m_i^\dagger
    m_j\rangle_{t=T}$ (circles) after the ramp, and those of the
    initial atomic state, $\langle
    a^\dagger_{i\downarrow}a^\dagger_{i\uparrow}
    a_{j\uparrow}a_{j\downarrow}\rangle_{t=0}$
    (solid line).  The plot for $U_{\uparrow\downarrow}=5J$ has been
    shifted up by $0.2$.}
\end{figure}

We have simulated numerically the ramping of small lattices, $L=10$ to
$32$ sites, with an initial number of atoms $N_{\uparrow,\downarrow} =
L/2, 3L/4$. The value of the molecular coupling has been fixed to
$\Omega = 1$ and the interaction has been ramped according to $E_m =
U_{\uparrow\downarrow} + 4 \Omega (1 - 2 t/T)$ using the ideal ramp
time $T=1.5/\Omega$. We have used two particular values of the
inter-species interaction, $U_{\uparrow,\downarrow}/\Omega = 0,2$, and
scanned different values of the hopping $J/\Omega\in [0,0.4]$. The
initial condition was always the ground state of the model with these
values of $U_{\uparrow\downarrow}/J$ and no coupling. These states
contain the correlations that we want to measure.

The main conclusions is that indeed the correlations of the molecules
are almost those of the initial state of the atoms (\ref{corr}), even
for $J=0.4\Omega$ when the process has not been adiabatic. An
intuitive explanation is that hopping is strongly suppressed as we
approach the resonance, due to the mixing between atomic states, which
can hop, and molecular states, which are slower. We can say that the
molecules thus pin the atoms and \textit{measure} them. This
explanation is supported by a perturbation analysis at $J\ll \Omega$,
where one finds that a small molecular contamination slows the atoms
on the lattice. This analysis breaks down, however, for $J \sim
\Omega,$ the regime in which the numerical simulations are required.

\textit{Conclusions}.- We have performed a rather exhaustive
comparison of different methods for simulating the evolution of big,
one-dimensional quantum systems
\cite{vidal04,verstraete04b,dmrg-evol,feiguin04} with two other
methods developed in this work. All methods are substantially affected
by truncation and rounding errors, and to fight the latter we must
choose large integration time-steps.  However, the only algorithm
which succeeds for large time-steps is an Arnoldi method developed in
this work.  Finally, this algorithm can be applied to problems with
long range interactions.

Using this algorithm, we have simulated the dynamics of cold atoms in
a 1D optical lattice when crossing a Feschbach resonance. The main
conclusion is that with rather fast ramp times it is possible to map
the correlations of the atomic cloud (two Tonks gases in this case)
and use this as a measuring tool in current experiments. This result
connects with similar theoretical predictions for fermions in
Ref.~\cite{other-papers}. Simple generalizations of this work will
allow us in the future to analyze losses and creation of entanglement
with the help of the molecular component.

\end{document}